\documentclass[]{WileyASNA-v1}

\usepackage[utf8]{inputenc}
\usepackage{calc}
\usepackage{anyfontsize}
\usepackage{hyperref}

\newlength{\okinalen}
\newcommand{\okina}{\hbox to.666\okinalen{\hss`\hss}}

\articletype{Article Type}%

\received{1 July 2020}
\revised{29 August 2020}
\accepted{18 September 2020}

\begin{document}

\title{Gaia Search for stellar Companions of TESS Objects of Interest}

\author[1]{M. Mugrauer*}

\author[1]{K.-U. Michel}

\authormark{Mugrauer \& Michel}

\address[1]{Astrophysikalisches Institut und Universit\"{a}ts-Sternwarte Jena}

\corres{M. Mugrauer, Astrophysikalisches Institut und Universit\"{a}ts-Sternwarte Jena, Schillerg\"{a}{\ss}chen 2, D-07745 Jena, Germany.\newline \email{markus@astro.uni-jena.de}}

\abstract{The first results of a new survey are reported, which explores the 2nd data release of the ESA-Gaia mission, in order to search for stellar companions of (Community) TESS Objects of Interest and to characterize their properties. In total, 193 binary and 15 hierarchical triple star systems are presented, detected among 1391 target stars, which are located at distances closer than about 500\,pc around the Sun. The companions and the targets are equidistant and share a common proper motion, as it is expected for gravitationally bound stellar systems, proven with their accurate Gaia astrometry. The companions exhibit masses in the range between about 0.08\,$M_\odot$ and 3\,$M_\odot$ and are most frequently found in the mass range between 0.13 and 0.6\,$M_{\odot}$. The companions are separated from the targets by about 40 up to 9900\,au, and their frequency continually decreases with increasing separation. While most of the detected companions are late K to mid M dwarfs, also 5 white dwarf companions were identified in this survey, whose true nature is revealed by their photometric properties.}


%%\fundingInfo{Funding info text.}

\maketitle

\section{Introduction}

A key aspect in the diversity of exoplanets is the multiplicity of their host stars. Stellar companions on close orbits with a few tens of astronomical units (au) but also wide companions with separations up to a few thousand au might significantly influence the formation process of planets in the gas and dust disks around their host stars, and/or the long-term evolution of their orbits \citep[see e.g.][]{kley2008,thebault2015,wu2007,kaib2013}. In order to detect such stellar systems with exoplanets and to characterize their properties, surveys are ongoing to search for stellar companions of exoplanet host stars, using seeing-limited, lucky-, and high contrast adaptive optics imaging, as well as catalogues searches \citep[see e.g.][]{roell2012, mugrauer2014, mugrauer2015, mugrauer2020}. In addition, recently \cite{mugrauer2019} explored the 2nd data release of the European Space Agency (ESA) Gaia mission \citep[Gaia DR2 from hereon, ][]{gaiadr2}, which provides manifold highly accurate astro- and photometric data of a huge number of objects, located across the whole sky, which make it an excellent database for the search of stellar companions of exoplanet host stars.

In the recent years the number of exoplanet host stars rapidly increased, which were mainly detected by space missions, launched to search for exoplanets by using the transit technique. Most exoplanets were detected so far by the Kepler mission \citep{borucki2010}, which monitored the photometry of hundreds of thousand stars in selected fields on the sky over about 9 years. In 2018, the year when the Kepler mission ended, the Transiting Exoplanet Survey Satellite \citep[TESS, ][]{ricker2015} was launched, which carries out photometric observations of 26 wide field sectors of the sky, each continually monitored by the satellite for about 27 days during the first two years of its mission. Thereby, TESS will observe more than 80\,\% of the whole sky providing data to search for transit signals in the light curves of millions of stars. By the end of May 2020 already more than 1800 stars with promising dips in their light curves, which could be caused by potential exoplanets, the so called TESS Objects of Interest (TOIs) were identified. Furthermore, in the light curves of about additional 300 stars, the so called Community TESS Objects of Interest (CTOIs), signatures of potential exoplanet candidates were identified in the TESS imaging data by the community, using different photometric pipelines. If the existence of these exoplanet candidates are confirmed by follow-up observations, which are currently ongoing, the associated (C)TOIs are newly identified exoplanet host stars.

The number of (C)TOIs is continuously growing during the successful execution of the TESS mission and the analysis of its photometric data. Therefore, a new survey was initiated at the Astrophysical Institute and University Observatory Jena, in order to explore the multiplicity of all these potential exoplanet host stars and to characterize the properties of detected companions by exploiting data from the Gaia DR2.

In the following section the project is described in detail, and its first results are presented in section 3. A summary of the current status of the survey, as well as an outlook of the project is presented in the last section of this paper.

\section{Search for stellar companions of (C)TOIs by exploring the Gaia DR2}

In the survey, presented here, stellar companions of the investigated (C)TOIs are identified at first as sources, which are located at the same distances as the targets, and secondly share a common proper motion with these stars. In order to clearly detect co-moving companions and prove that these objects and the (C)TOIs are equidistant, only sources are taken into account in this survey, which are listed in the Gaia DR2 with accurate five parameter astrometric solutions ($\alpha$, $\delta$, $\pi$, $\mu_{\alpha}cos(\delta)$, $\mu_{\delta}$), and exhibit significant measurements of their parallaxes ($\pi/\sigma(\pi) > 3$) and proper motions ($\mu/\sigma(\mu) > 3$). Thereby sources, listed with a negative parallax, are neglected. As a parallax uncertainty of 0.7\,mas is reached for faint sources down to $G = 20$\,mag in the Gaia DR2, the survey is furthermore constrained to (C)TOIs, which are located within a distance of 500\,pc around the Sun (i.e. $\pi > 2$\,mas), to assure $\pi/\sigma(\pi) > 3$ even for the faintest companions, detectable in this survey. This distance constraint is slightly relaxed to $\pi + 3\sigma(\pi)>2$\,mas, i.e. taking into account also the parallax uncertainty of the (C)TOIs.

By the end of May 2020, in total 1391 stars are listed in the (C)TOI Release of the \verb"Exoplanet" \verb"Follow-up" \verb"Observing" \verb"Program" for TESS (ExoFOP-TESS)\footnote{Online available at:\newline\url{https://exofop.ipac.caltech.edu/tess/view_toi.php}\newline\url{https://exofop.ipac.caltech.edu/tess/view_ctoi.php}}, which fulfil this distance constraint, and are therefore selected as targets for the survey, presented here. Thereby, (C)TOIs with dips in their light curves, which could be already classified as false positive detections by follow-up observations, carried out in the course of the ExoFOP-TESS, were removed from the target list.

\begin{figure*}
\resizebox{\hsize}{!}{\includegraphics{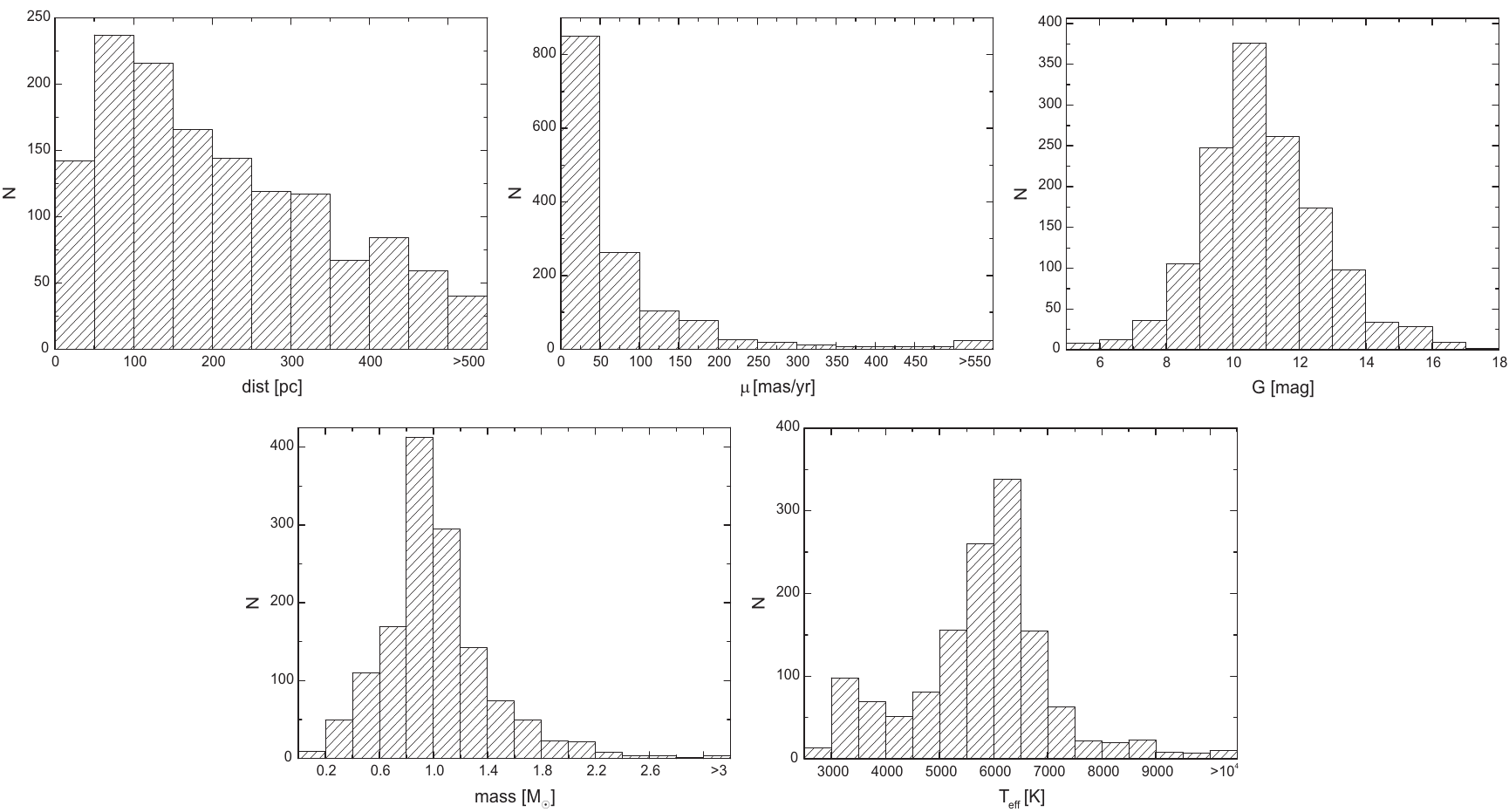}}\caption{The histograms of the individual properties of all targets.}\label{HIST_TARGETS}
\end{figure*}

The histograms of the properties of all targets are summarized in Fig.\,\ref{HIST_TARGETS}. The distances ($dist$) and the total proper motions ($\mu$) of the targets are derived with their accurate Gaia DR2 parallaxes ($dist[pc]=1000/\pi[mas]$) and proper motions in right ascension and declination. The G-band magnitudes of all targets are taken from the Gaia DR2, their masses and effective temperatures ($T_{eff}$) from the Starhorse Catalog \citep[SHC from hereon,][]{anders2019}, respectively.

The targets of this survey are located at distances between 7 up to about 550\,pc and exhibit proper motions in the range between about 1 up to 2100\,mas/yr, G-band magnitudes from 5.2 to 17.2\,mag, effective temperatures from 2700 up to 14500\,K, and masses, which range between about 0.16 and 5\,$M_{\odot}$. According to the cumulative distribution functions of the individual properties, the targets are most frequently located at distances between about 30 and 140\,pc, and exhibit proper motions in the range between about 5 and 20\,mas/yr, as well as G-band magnitudes from $G=9.8$ to 11.5\,mag. The targets are mainly solar like stars with masses in the range between 0.8 and 1.1\,$M_{\odot}$. This population also emerges in the $T_{eff}$ distribution of the targets at intermediate temperatures of about 6000\,K. In addition, another but fainter pile-up of targets is evident in this distribution at lower effective temperatures between about 3000 and 3700\,K, which is the early to mid M dwarf population.

The angular search radius around the selected targets, within which the companion search is carried out, is limited to $r [\text{arcsec}] = 10 \pi[\text{mas}]$, with $\pi$ the Gaia DR2 parallaxes of the (C)TOIs. This allows the detection of companions with projected separations up to 10000\,au around the stars, which guarantees an effective companion search, that will detect the vast majority of all wide companions of the targets, as described by \cite{mugrauer2019}.

All sources with an accurate five parameter astrometric solution, listed in the Gaia DR2, which are located within the used search radius around the targets are considered as companion candidates. In total, 78572 such objects were detected around 1170 targets, investigated in the course of this survey. The companionship of all these candidates was tested based on their accurate Gaia DR2 astrometry and that of the associated (C)TOIs, exactly following the procedure, described in \cite{mugrauer2019}. The vast majority of these sources\linebreak ($>99.7$\,\%) could be excluded as companions, as they are either not located at the same distances as the (C)TOIs and/or do not share a common proper motion with these stars, i.e. their parallaxes and proper motions significantly differ from each other. In contrast, for 221 candidates, which are presented in this paper, their companionship to the (C)TOIs could clearly be proven with their accurate Gaia DR2 astrometry. The properties of these (C)TOIs and their detected companions are described in detail in the next section.

\section{(C)TOIs and their detected stellar companions}

\begin{figure}
\resizebox{\hsize}{!}{\includegraphics{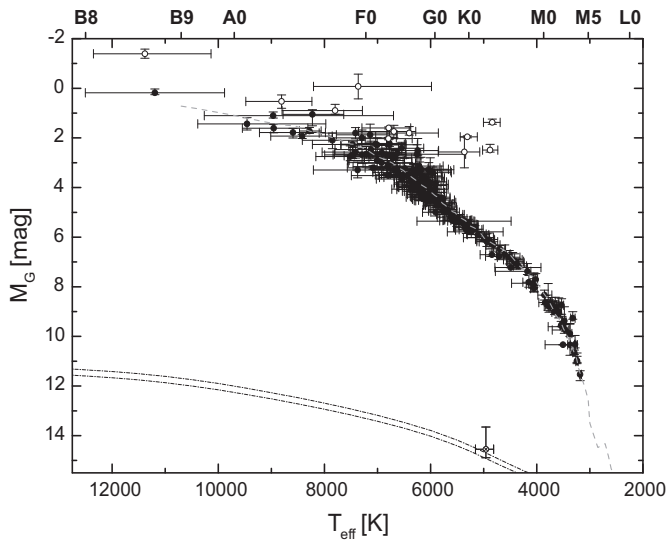}}\caption{The $T_{eff}$-$M_G$ diagram of all (C)TOIs with detected companions, presented here. The main-sequence is shown as grey dashed line, the evolutionary tracks of DA dwarfs with masses of 0.5 and 0.6\,$M_\odot$ as black dash-dotted lines, respectively. (C)TOIs, listed in the SHC with surface gravities $log(g[cm/s^{-2}]])\leq3.8$, are illustrated as white circles, those with larger surface gravities with black circles, respectively. The white dwarf TOI\,1690\,C is plotted as a crossed circle.}\label{HRDCTOIS}
\end{figure}

The masses, effective temperatures, and absolute G-band magnitudes of the (C)TOIs with detected companions, presented here, are all listed in the SHC, except for TOI\,1690. This target is identified in our survey as the tertiary component of a hierarchical triple star system and therefore is named as TOI\,1690\,C. The absolute magnitude of this star was derived as described below for the detected companions and we use here the Apsis-Priam temperature estimate of the star to plot it together with the other (C)TOIs with detected companions in the $T_{eff}$-$M_G$ diagram, which is shown in Fig.\,\ref{HRDCTOIS}. For comparison we plot in this diagram the main-sequence from \cite{pecaut2013}\footnote{Online available at: \url{http://www.pas.rochester.edu/~emamajek/EEM_dwarf_UBVIJHK_colors_Teff.txt}}, as well as evolutionary mass tracks of DA white dwarfs, calculated by the white dwarf models from \cite{holberg2006}, \cite{kowalski2006}, \cite{tremblay2011}, and \cite{bergeron2011}.

The vast majority of all targets with detected companions are main-sequence stars. In contrast, TOI\,1690\,C is clearly located below the main-sequence but its photometry is fully consistent with that expected for DA white dwarfs. Furthermore, few (C)TOIs are located significantly above the main-sequence and all these stars exhibit surface gravities $log(g[cm/s^{-2}]])\leq3.8$, as listed in the SHC, hence are classified as giants.

The parallaxes, proper motions, apparent G-band magnitudes, and extinction estimates of the (C)TOIs and their companions, detected in this survey, are summarized in Tab.\,\ref{TAB_COMP_ASTROPHOTO}, which lists in total, 193 binary, and 15 hierarchical triple star systems.

For each detected companion we derived its angular separation ($\rho$) and position angle ($PA$) to the associated (C)TOI, using the accurate Gaia DR2 astrometry of the individual objects. The obtained relative astrometry of the companions is summarized in Tab.\,\ref{TAB_COMP_RELASTRO}, which lists also its uncertainty, which is below about 1\,mas in angular separation, and 0.05\,$^{\circ}$ in position angle, respectively.

The difference $\Delta \pi$ between the parallaxes of the (C)TOIs and their companions together with its significance $sig\text{-}\Delta\pi$ was determined (in addition also by taking into account the astrometric excess noise of the individual objects) and is summarized in Tab.\,\ref{TAB_COMP_RELASTRO}. In the same table we list for each companion its differential proper motion $\mu_{rel}$ relative to the associated (C)TOI with its significance, as well as its $cpm$-$index$\footnote{The degree of common proper motion of a detected companion with the associated (C)TOI is characterized by its common proper motion (cpm) index, which is defined by \cite{mugrauer2019} as: $cpm\text{-}index = | \overrightarrow{\mu}_{(C)TOI} + \overrightarrow{\mu}_{Comp}| / \mu_{rel}$\linebreak with $\overrightarrow{\mu}_{(C)TOI}$ the proper motion of the (C)TOI, and $\overrightarrow{\mu}_{Comp}$ the proper motion of the companion, as well as its differential proper motion $\mu_{rel}=| \overrightarrow{\mu}_{(C)TOI} - \overrightarrow{\mu}_{Comp}|$.}.

The parallaxes of the individual components of the stellar systems, presented here, do not significantly differ from each other ($sig\text{-}\Delta\pi \leqq 3$), when the astrometric excess noise is taken into account. This clearly proves the equidistance of the detected companions with the (C)TOIs, as expected for components of physically associated stellar systems. Furthermore, the vast majority of the detected companions (more than 90\,\% of all) exhibit a $cpm\text{-}index> 10$, and all companions reach a $cpm\text{-}index > 3$. Hence, the detected companions and the associated (C)TOIs clearly form common proper motion pairs, as expected for gravitationally bound stellar systems.
\begin{figure}
\resizebox{\hsize}{!}{\includegraphics{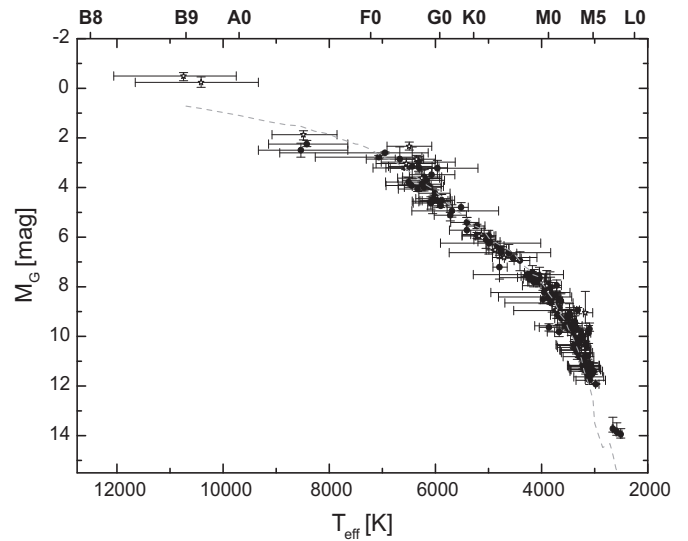}}\caption{This $T_{eff}$-$M_G$ diagram shows all detected companions, whose effective temperatures are either listed in the SHC, or for which Apsis-Priam temperature estimates are available in the Gaia DR2. Companions, which are the primary components of their stellar systems, are illustrated as star symbols. The main-sequence is plotted as dashed grey line for comparison.}\label{HRDCOMPS}
\end{figure}

\begin{table*}[h!]
\caption{The photometry of the five white dwarf companions, detected in this survey. For each companion we list the color-difference $\Delta(B_P-R_P)$, and the G-band magnitude-difference $\Delta G$ to the associated (C)TOI, its apparent $(B_P-R_P)$ color, as well as its derived intrinsic color $(B_{P}-R_{P})_{0}$, and effective temperature $T_{eff}$.}\label{TAB_WDS_PROPS}
\centering
\begin{tabular}{lccccc}
\hline
Companion         & $\Delta(B_{P}-R_{P})$ $[$mag$]$ & $\Delta G$ $[$mag$]$        & $\left(B_{P}-R_{P}\right)$ $[$mag$]$      & $(B_{P}-R_{P})_{0}$ $[$mag$]$  & $T_{eff}$$[$K$]$\\
\hline
TOI\,1259\,B      & $-0.5980\pm0.0782$    & $7.4165\pm0.0042$ & $0.7044\pm0.0781$  & $0.5984_{-0.1991}^{+0.1253}$   & $6473_{-419}^{+672}$\\
TOI\,1624\,B      & $-0.3844\pm0.0870$    & $9.0745\pm0.0046$ & $0.2896\pm0.0870$  & $0.2113_{-0.0940}^{+0.0949}$   & $7988_{-416}^{+477}$\\
TOI\,249\,C       & $-1.0049\pm0.0117$    & $5.8888\pm0.0010$ & $0.6967\pm0.0116$  & $0.4927_{-0.1216}^{+0.0795}$   & $6801_{-244}^{+444}$\\
CTOI\,53309262\,B & $-2.4847\pm0.0125$    & $2.1228\pm0.0015$ & $0.1275\pm0.0114$  & $-0.1167_{-0.1894}^{+0.0820}$  & $10670_{-994}^{+1879}$\\
TOI\,1703\,B      & $-0.7597\pm0.2914$    & $7.1962\pm0.0058$ & $0.1484\pm0.2914$  & $0.0044_{-0.2968}^{+0.2943}$   & $9269_{-1657}^{+1670}$\\
\hline
\end{tabular}
\end{table*}

The equatorial coordinates, as well as the derived absolute G-band magnitudes, projected separations, masses, and effective temperatures of all detected companions are summarized in Tab.\,\ref{TAB_COMP_PROPS}.

The absolute G-band magnitudes of the companions are taken from the SHC if available, or were derived with their apparent G-band photometry, the parallaxes of the (C)TOIs, as well as the Apsis-Priam G-band extinction estimates, listed in the Gaia DR2. Thereby, always the extinction estimates of the companions if available, otherwise those of the (C)TOIs were used. For systems with no G-band extinction estimates, listed for any of their components, extinction measurements of the (C)TOIs in the V-band, listed in the Vizier database \citep{ochsenbein2000}\footnote{Online available at: \url{http://vizier.u-strasbg.fr/}}, were used to derive the average and standard deviation of the V-band extinctions of these systems. These extinctions were transformed to the G-band using the relation $A_G/A_V=0.77$, determined by \cite{mugrauer2019}.

The projected separations of all companions were derived from their angular separations to the associated (C)TOIs and the parallaxes of these stars. For this purpose, we always used the parallaxes of the (C)TOIs, since they are usually more accurately determined than the parallaxes of the companions.

The masses and effective temperatures of all detected companions, presented here, including their uncertainties, are taken from the SHC if available, which applies to about 73\,\% of all companions. In Fig.\,\ref{HRDCOMPS} we plot these companions in a $T_{eff}$-$M_G$ diagram, together with the companions for which Apsis-Priam estimates of their effective temperatures are available\footnote{As recommended by \cite{andrae2018}, in this survey we use Apsis-Priam temperature estimates only if their flags are equal to $\texttt{1A000E}$ with $\texttt{A}$ and $\texttt{E}$ that can have any value.}, indicated by the $\texttt{PRI}$ flag in Tab.\,\ref{TAB_COMP_PROPS}. Except for the two brightest and hottest companions, which are located above the main sequence and exhibit low surface gravities ($log(g[cm/s^{-2}]])<3.8$), i.e. these companions are giants, the photometry of the majority of all detected companions is well consistent with that expected for main-sequence stars.

For the remaining 59 companions, whose properties are not listed in the SHC, we derived their masses and effective temperatures from their absolute G-band magnitudes via interpolation (indicated with the flag $\texttt{inter}$ in Tab.\,\ref{TAB_COMP_PROPS}) using the $M_G$-mass and $M_G$-$T_{eff}$ relations from \cite{pecaut2013}, adopting that these companions are main-sequence stars. In order to test this assumption, we compared the obtained effective temperatures of the companions with either their Apsis-Priam temperature estimates if available, or with the effective temperatures of the companions, derived from their $(B_P-R_P)$ colors and Apsis-Priam reddening estimates $E(B_P-R_P)$ or if not available those of the associated (C)TOIs, using the $(B_P-R_P)_0$-$T_{eff}$ relation from \cite{pecaut2013}.

For all but five of these companions their effective temperatures, derived from their absolute magnitudes by assuming that they are main-sequence stars, agree well with either their Apsis-Priam temperature estimates or the temperatures, obtained from their colors. The average deviation of the different temperature estimates is about 300\,K, well consistent with the average uncertainty of the derived effective temperatures. Hence, we conclude that these companions are all main-sequence stars.

Also the $(B_P-R_P)$ colors of the (C)TOIs and their companions are compared with each other, indicated by the \verb"BPRP" flag in Tab.\,\ref{TAB_COMP_PROPS}, if a color-comparison is possible. For main-sequence stars we expect that companions, which are fainter/brighter than the (C)TOIs, appear redder/bluer than the stars, and this holds for the majority of all detected companions except for the five stars, discussed below in more detail.

\begin{figure}
\resizebox{\hsize}{!}{\includegraphics{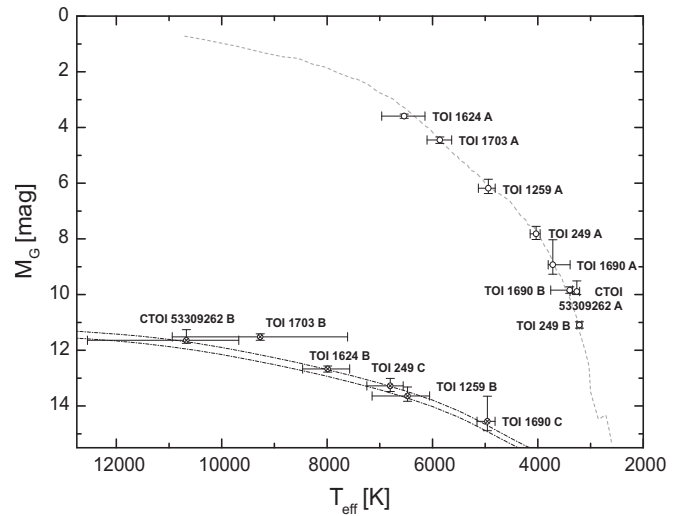}}\caption{$T_{eff}$-$M_G$ diagram of the stellar systems with white dwarf components, detected in this survey. The main-sequence is plotted as grey dashed line, and the evolutionary mass tracks of DA white dwarfs with masses of 0.5 and 0.6\,$M_{\odot}$ as black dash-dotted lines, respectively.}\label{HRD_WDS}
\end{figure}

The detected companions TOI\,249\,C, TOI\,1259\,B, TOI\,1624\,B, TOI\,1703\,B, and CTOI\,53309262\,B are all several magnitudes fainter than the associated (C)TOIs, but appear significantly bluer than their primaries. The temperatures of these companions, derived from their absolute G-band magnitudes, adopting that they are main-sequence stars, is about 3500 to 7500\,K lower than the temperatures, obtained from their colors. We summarize the properties of these companions in Tab.\,\ref{TAB_WDS_PROPS} and plot them together with the other components, detected in these stellar systems, in a $T_{eff}$-$M_G$ diagram, which is shown in Fig.\,\ref{HRD_WDS}.

\begin{figure*}
\resizebox{\hsize}{!}{\includegraphics{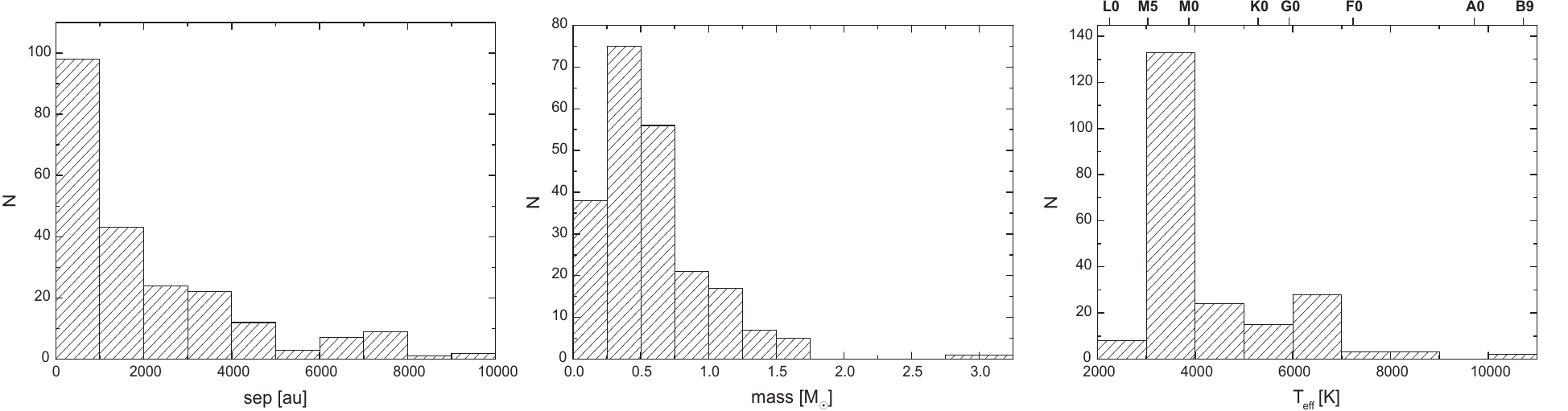}}\caption{The histograms of the properties of the companions, detected in this survey.}\label{HIST_COMPS}
\end{figure*}

While the brighter components of these systems are all main-sequence stars the five faint companions are clearly located below the main-sequence but their photometry is consistent with that expected for white dwarfs. Hence, we conclude that TOI\,249\,C, TOI\,1259\,B, TOI\,1624\,B, TOI\,1703\,B, and CTOI\,53309262\,B are white dwarf companions of the associated (C)TOIs, which is indicated with the $\texttt{WD}$ flag in Tab.\,\ref{TAB_COMP_PROPS}. Follow-up spectroscopic observations are needed to further characterize the properties of these degenerated companions.

The histograms of the properties of the companions, presented here, are illustrated in Fig.\,\ref{HIST_COMPS}. The companions exhibit angular separations to the (C)TOIs, in the range between about 0.8 and 160\,arcsec, which corresponds to projected separations of about 42 up to 9865\,au. According to the underlying cumulative distribution function, the frequency of the companions continually decreases with increasing projected separation and half of all companions exhibit projected separations of less than 1300\,au. In total, 9 stellar systems (all binaries) are detected with project separations below 100\,au, namely: TOI\,253\,AB, TOI\,1215\,BA, TOI\,1450\,AB, TOI\,1452\,AB, TOI\,1634\,AB, TOI\,1746\,AB, CTOI\,293689267\,BA, CTOI\,327667965\,AB, and CTOI\,350190639\,AB, i.e. these systems are the most challenging environments for planet formation, identified in this study.

The masses of the companions range form the substellar/stellar mass border at about 0.08\,$M_{\odot}$ up to $\sim$\,3\,$M_{\odot}$ (average mass is about 0.6\,$M_\odot$). The highest companion frequency is found in the cumulative distribution function in the mass range between 0.13 and 0.6\,$M_{\odot}$, which corresponds beside detected white dwarf companions mainly to mid M to late K dwarfs, according to the relation between mass and spectral type (SpT), described by \cite{pecaut2013}. For higher masses the companion frequency continually decreases. This peak in the companions population is also detected in the distribution of their effective temperatures, which exhibits the highest frequency of companions in the temperature range between 3000 and 4000\,K. In this distribution also a second but fainter pile-up of companions is prominent, which is located between 5900 and 6600\,K and corresponds to solar like stars with SpTs of G0 to F4, according to the $T_{eff}$-$\text{SpT}$ relation from \cite{pecaut2013}.

\begin{figure}[h]
\resizebox{\hsize}{!}{\includegraphics{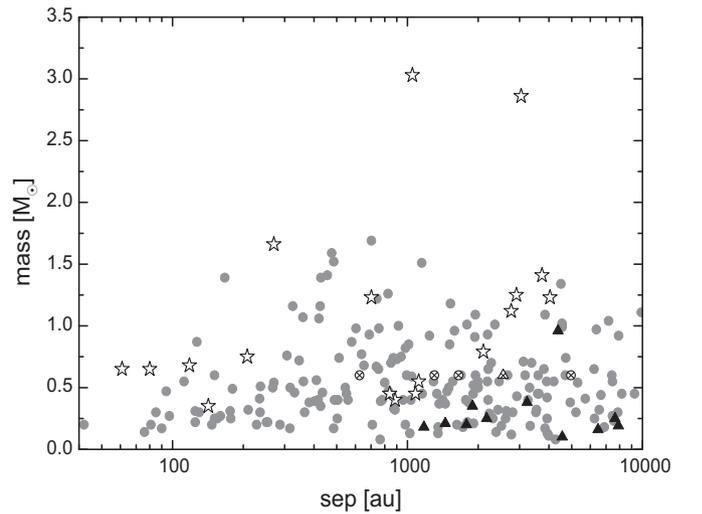}}\caption{The separation-mass diagram of the companions, detected in this survey. Companions, which are the primary components of their stellar systems, are plotted as star symbols, those which are secondaries as circles and tertiary components as triangles, respectively. Detected white dwarf companions, for which a mass of 0.6\,$M_{\odot}$ is adopted, are plotted with white crossed symbols.}\label{SEPMASS}
\end{figure}

The detected companions are usually the fainter and lower-mass secondary or tertiary components in their stellar systems, as illustrated in the separation-mass diagram of the detected companions, which is shown in Fig.\,\ref{SEPMASS}. Among all 221 companions, presented here, 18 are the primary, 191 the secondary, and 12 the tertiary components of their stellar systems.\newpage

\begin{figure}[h]
\resizebox{\hsize}{!}{\includegraphics{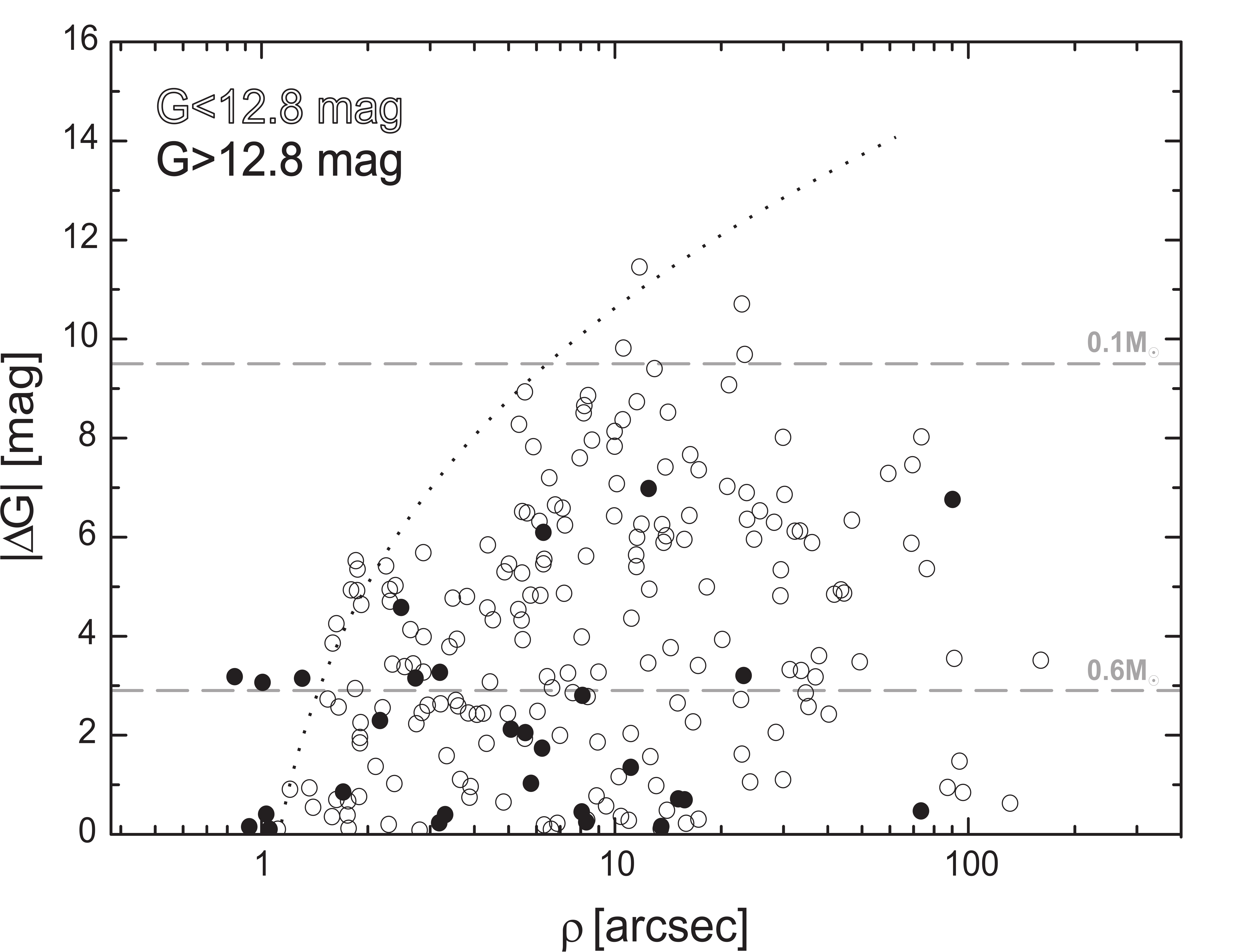}}\caption{The magnitude-differences of all detected companions plotted versus their angular separations to the associated (C)TOIs. The Gaia detection limit, derived by \cite{mugrauer2019}, is shown as dotted grey line for comparison. The expected average magnitude-difference for companions with 0.1 or 0.6\,$M_{\odot}$ is drawn as grey dashed horizontal lines. Companions of (C)TOIs brighter than $G=12.8$\,mag are plotted as open circles those of (C)TOIs, which are fainter than that magnitude limit, as filled black circles, respectively.}\label{LIMIT}
\end{figure}

In order to characterize the detection limit, reached in this survey, we plot the magnitude-differences of all detected companions over their angular separations to the associated (C)TOIs, as shown in Fig.\,\ref{LIMIT}.

In stellar systems with primary stars brighter than $G=12.8$\,mag (about 86\% of the targets of this survey) companions are detectable at angular separations larger than about 1\,arcsec, very well consistent with the limit found by \cite{mugrauer2019} among exoplanet host stars. In this multiplicity survey there is one companion reported at an angular separation of 1.3\,arcsec, which exhibits a magnitude difference of 4.2\,mag. Here we detected two companions at angular separations slightly below 1\,arcsec with magnitude-differences of about 3\,mag, which all exceed the given detection limit. However, as illustrated in Fig.\,\ref{LIMIT} all these stars are companions of faint primaries with G-band magnitudes $G>12.8$\,mag. Hence, for such faint targets companions with magnitude-differences up to 3\,mag are detectable with Gaia even slightly below the 1\,arcsec separation limit. The expected magnitude-differences between the targets of this survey and low-mass main-sequence companions (indicated with grey dashed lines in Fig.\,\ref{LIMIT}) are estimated with the expected absolute G-band magnitudes of these stars, as listed by \cite{pecaut2013}, and the average absolute G-band magnitude of our targets ($M_G=4.9$\,mag). As shown in Fig.\,\ref{LIMIT} a magnitude difference of $\sim$\,4\,mag is reached at an angular separation of about 1.5\,arcsec around the targets of this survey. This allows the detection of companions with masses down to about 0.6\,$M_\odot$ (average mass of all detected companions) which are separated from the (C)TOIs by more than 315\,au. Furthermore, companions with masses down to $\sim$\,0.1\,$M_\odot$ are detectable beyond about 6\,arcsec, which corresponds to a projected separation of 1260\,au at the average target distance of 210\,pc.

\section{Summary and Outlook}

The goal of the survey, which is presented here, is the detection and characterization of stellar companions of (C)TOIs, i.e. of potential exoplanet host stars. By the end of May 2020 the multiplicity of 1391 (C)TOIs could already be explored in the course of this survey using data from the Gaia DR2 and co-moving companions were detected around 208 targets. Beside 193 binaries, whose properties are described here, also 15 hierarchical triple star systems were detected, in which either a (C)TOI exhibits a close and a wide companion, or a close binary companion instead, which is located at a wider angular separation.

As it is expected for the components of gravitationally bound stellar systems the (C)TOIs and the detected companions are equidistant and share a common proper motion, as proven with their accurate Gaia DR2 parallaxes and proper motions. In particular, the direct proof of equidistance of the individual components of the stellar systems, as done in this survey by comparing their parallaxes, was not feasible in earlier multiplicity surveys as the targets and/or their companions are not detected by the ESA-Hipparcos mission \citep{perryman1997}. However, 84 companions, identified in this survey, are already listed in the  Washington Double Star Catalog \citep[WDS from hereon,][]{mason2001}, either as co-moving companions, or as companion candidates of the (C)TOIs, which still need confirmation of their companionship, eventually yielded by this survey. Although the WDS is currently the largest and most complete available catalogue of multiple star systems, which contains relative astrometric measurements of these systems over a period of more than 300 years, in this study 137 (i.e. 62\,\% of all) companions were detected, which are not listed in the WDS, indicated with the $\bigstar$ flag in the last column of Tab.\,\ref{TAB_COMP_RELASTRO}. This demonstrates the great potential of the ESA-Gaia mission for multiplicity studies of stars, in particular for the detection of wide companions, as it is illustrated with the derived detection limit of this survey, shown in Fig.\,\ref{LIMIT}.

\begin{figure*}
\begin{center}\includegraphics[width=15cm]{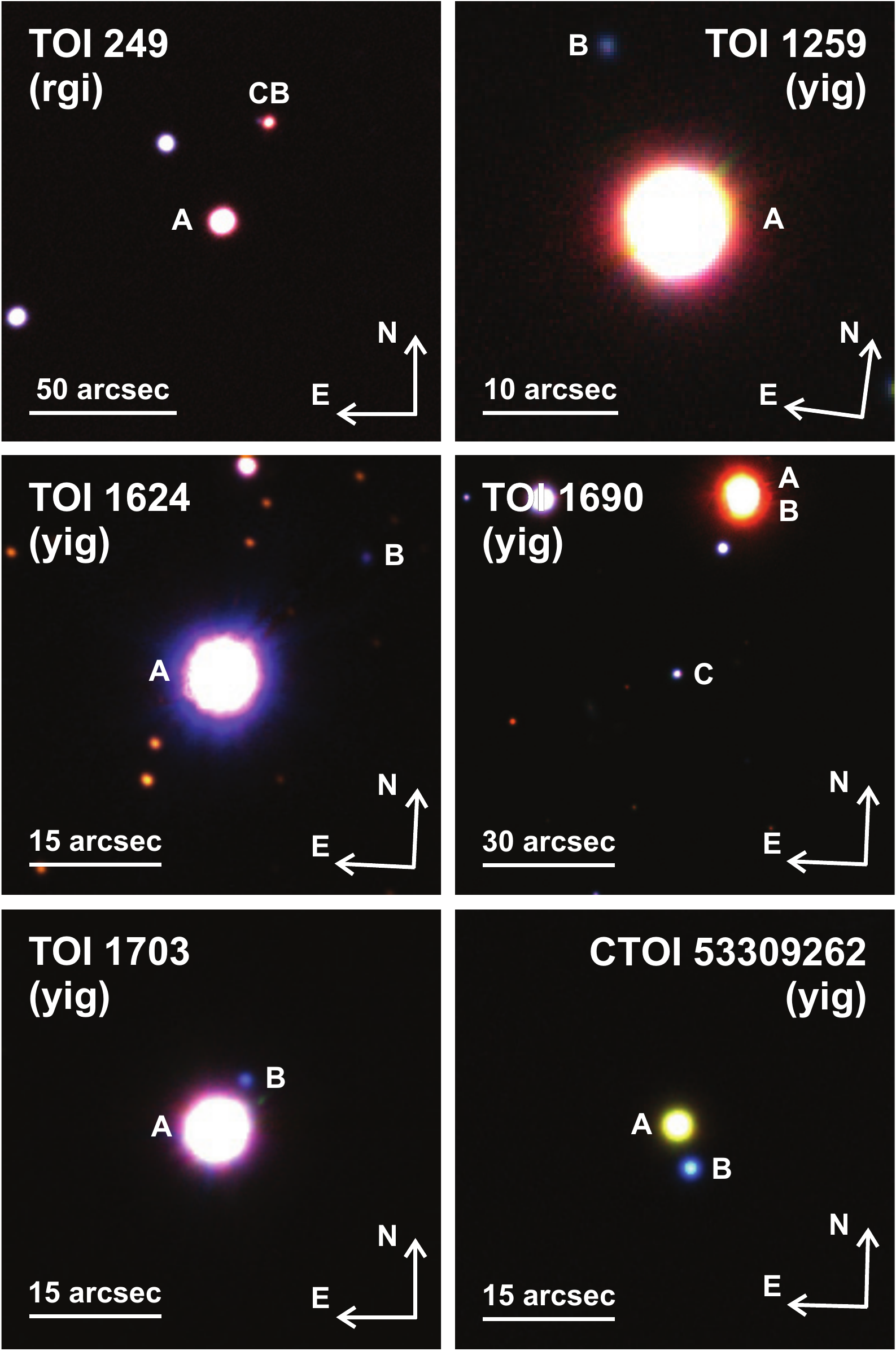}\caption{Color(RGB)-composit images of all stellar systems with white dwarf components, detected in this survey. The image of TOI\,249 is created from imaging data, taken in the course of the SkyMapper southern sky survey in the r-, g-, and i-band. The images of TOI\,1259, TOI\,1624, TOI\,1690, TOI\,1703, and CTOI\,53309262 are color-composites, made of y-, i-, and g-band images, taken by the Panoramic Survey Telescope and Rapid Response System (Pan-STARRS).}\label{PICS}\end{center}
\end{figure*}

On average, all stellar companions with masses down to about 0.1\,$M_\odot$ are detectable in this study around the targets beyond $\sim$\,6\,arcsec (or 1260\,au of projected separation), and approximately half of all detected companions exhibit such separations. In total, companions are identified with projected separations between about 40 and 9900\,au and the frequency of companions continually decreases with increasing projected separation. The companions, detected in this survey, exhibit masses in the range between the substellar/stellar mass border at about 0.08\,$M_\odot$ and 3\,$M_\odot$, most frequently found in the mass range between 0.13 and 0.6\,$M_\odot$. Beside low-mass main sequence stars (mainly late K to mid M dwarfs) also 5 white dwarfs could be identified as co-moving companions of the (C)TOIs, whose true nature was revealed in this survey, using their accurate astro- and photometric properties, as listed in the Gaia DR2. Color-composite images of all evolved stellar systems with white dwarf components, identified in this survey, are shown in Fig.\,\ref{PICS}.

Beside the 221 companions presented here we have also detected companions around 54 additional targets, which are not described in this paper, as the associated (C)TOIs are all known exoplanet host stars, listed in the \verb"Extrasolar Planets Encyclopaedia" \citep{schneider2011}\footnote{Available online at: \url{http://exoplanet.eu/}}. The companions of these targets were already characterized in \cite{mugrauer2019} or will be presented by \cite{michel2020}, respectively\footnote{The companions of TOI\,106, 107, 110, 123, 143, 185, 191, 229, 241, 265, 368, 398, 404, 418, 473, 479, 489, 490, 505, 511, 567, 675, 747, 752, 774 ,834, 1150, 1161, 1236, 1237, 1300, 1388, 1419, 1458, 1599, 1720, 1766, 1771, 1773, 1809, and 1909 are listed already in \cite{mugrauer2019}, and those of TOI\, 132, 174, 200, 396, 448, 732, 905, 966, 1067, 1148, 1165, 1924, and 1936 will be presented by \cite{michel2020}, respectively.}. Taking these additional companions into account, the current multiplicity rate of the (C)TOIs is at least about $19\pm1$\,\%, which is consistent with the minimum multiplicity rate of exoplanet host stars of $15\pm1$\,\%, recently determined by \cite{mugrauer2019}, especially considering that for some (C)TOIs the transit-like signals in their light curves could turn out to be false positive detections.

For 175 (i.e. about 79\,\% of all) companions, presented here, significant ($sig\text{-}\mu_{rel} \geq 3$) differential proper motions $\mu_{rel}$ relative to the associated (C)TOIs were detected. We derived the escape velocities $\mu_{esc}$ of all these companions using the approximation, described in \cite{mugrauer2019}. The differential proper motion of most of these companions is consistent with orbital motion. In contrast for 34 companions, their differential proper motions significantly exceed the expected escape velocities, indicating an increased degree of multiplicity, as discussed in \cite{mugrauer2019}. Indeed, 8 of these companions are located in already confirmed or potential hierarchical triple star systems but follow-up high contrast imaging observations are needed to further explore the multiplicity status of all these particular systems and their companions, which are listed in Tab.\,\ref{table_triples}.

\begin{table}[h!] \caption{List of all detected companions (sorted by their identifier), whose differential proper motions $\mu_{rel}$ relative to the (C)TOIs significantly exceed the expected escape velocities $\mu_{esc}$. Companions, which are already known to be members of hierarchical triple star systems, are indicated with $\bigstar\bigstar\bigstar$, and those in potential hierarchical triple star systems with ($\bigstar\bigstar\bigstar$), respectively.}
\begin{center}
\begin{tabular}{lccc}
\hline
Companion          & $\mu_{rel}$ [mas/yr] & $\mu_{esc}$ [mas/yr] &\\
\hline
TOI\,129\,C        & $5.17\pm0.32 $       & $2.47\pm0.05 $ &$\bigstar\bigstar\bigstar$\\
TOI\,179\,BC       & $6.05\pm0.35 $       & $4.89\pm0.08 $ &$\bigstar\bigstar\bigstar$\\
TOI\,330\,B        & $5.05\pm0.37 $       & $2.02\pm0.07 $ &\\
TOI\,422\,A        & $5.74\pm0.06 $       & $1.81\pm0.09 $ &($\bigstar\bigstar\bigstar$)\\
TOI\,510\,B        & $7.21\pm0.06 $       & $5.53\pm0.23 $ &\\
TOI\,658\,B        & $5.57\pm0.24 $       & $2.31\pm0.13 $ &\\
TOI\,721\,B        & $2.18\pm0.06 $       & $0.83\pm0.03 $ &\\
TOI\,811\,B        & $1.53\pm0.05 $       & $1.25\pm0.01 $ &\\
TOI\,833\,B        & $8.96\pm0.18 $       & $5.73\pm0.09 $ &$\bigstar\bigstar\bigstar$\\
TOI\,858\,A        & $2.86\pm0.18 $       & $0.99\pm0.05 $ &\\
TOI\,866\,B        & $0.62\pm0.06 $       & $0.31\pm0.02 $ &\\
TOI\,1099\,B       & $37.03\pm0.18\,\,\,$ & $31.88\pm0.97\,\,\,$ &\\
TOI\,1145\,A       & $1.43\pm0.13 $       & $0.86\pm0.05 $ &\\
TOI\,1254\,B       & $3.92\pm0.16 $       & $1.66\pm0.09 $ &\\
TOI\,1310\,B       & $2.55\pm0.18 $       & $1.03\pm0.05 $ &\\
TOI\,1315\,B       & $1.06\pm0.07 $       & $0.55\pm0.01 $ &\\
TOI\,1520\,B       & $2.27\pm0.12 $       & $0.58\pm0.03 $ &\\
TOI\,1557\,B       & $1.84\pm0.18 $       & $0.70\pm0.05 $ &\\
TOI\,1671\,B       & $1.65\pm0.07 $       & $0.72\pm0.03 $ &\\
TOI\,1690\,A       & $15.36\pm0.15\,\,\,$ & $11.19\pm0.11\,\,\,$ &$\bigstar\bigstar\bigstar$\\
TOI\,1709\,B       & $2.24\pm0.13 $       & $1.08\pm0.08 $ &\\
TOI\,1733\,B       & $1.38\pm0.19 $       & $0.52\pm0.03 $ &$\bigstar\bigstar\bigstar$\\
TOI\,1733\,C       & $1.54\pm0.27 $       & $0.46\pm0.03 $ &$\bigstar\bigstar\bigstar$\\
TOI\,1749\,B       & $11.16\pm1.20\,\,\,$ & $1.66\pm0.01 $ &\\
TOI\,1831\,B       & $4.77\pm0.21 $       & $2.47\pm0.10 $ &\\
TOI\,1846\,B       & $13.71\pm1.73\,\,\,$ & $2.09\pm0.03 $ &\\
TOI\,1855\,B       & $3.41\pm0.09 $       & $1.33\pm0.05 $ &\\
TOI\,1857\,B       & $1.20\pm0.08 $       & $0.68\pm0.03 $ &\\
TOI\,1859\,B       & $3.81\pm0.91 $       & $0.94\pm0.06 $ &\\
{\fontsize{7}{0}\selectfont CTOI\,146129309\,A} & $1.51\pm0.07 $       & $0.60\pm0.03 $ &\\
{\fontsize{7}{0}\selectfont CTOI\,207080350\,B} & $1.83\pm0.09 $       & $1.37\pm0.06 $ &\\
{\fontsize{7}{0}\selectfont CTOI\,381854774\,C} & $1.07\pm0.10 $       & $0.58\pm0.03 $ &$\bigstar\bigstar\bigstar$\\
{\fontsize{7}{0}\selectfont CTOI\,404927661\,B} & $1.56\pm0.22 $       & $0.67\pm0.03 $ &\\
{\fontsize{7}{0}\selectfont CTOI\,441422527\,B} & $5.19\pm0.11 $       & $4.83\pm0.03 $ &\\
\hline
\end{tabular}
\end{center}
\label{table_triples}
\end{table}

The survey, whose first results are presented here, is an ongoing project, and its target list is steadily growing due to the continuing analysis of photometric data, collected by the TESS mission. The multiplicity of all these newly revealed (C)TOIs will be explored in the course of this survey and detected companions and their determined properties will be reported regularly in this journal and will also be made available online in the \verb"VizieR" database. Furthermore, there are many objects, listed in the Gaia DR2, which still lack a five parameter astrometric solution. Hence, there should exist further companions of the targets, investigated here, whose companionship can be proven with accurate astrometric measurements, provided by future data releases of the ESA-Gaia mission. The results of our survey combined with those of high-contrast imaging observations of the (C)TOIs, which can detect close companions with projected separations down to only a few au, will eventually provide a complete understanding of the multiplicity of all these potential exoplanet host stars.

\section*{Acknowledgments}

We made use of data from:

(1) the \verb"Simbad" and \verb"VizieR" databases operated at CDS in Strasbourg, France.

(2) the European Space Agency (ESA) mission Gaia (\url{https://www.cosmos.esa.int/gaia}), processed by the Gaia Data Processing and Analysis Consortium (DPAC, \url{https://www.cosmos.esa.int/web/gaia/dpac/consortium}). Funding for the DPAC has been provided by national institutions, in particular the institutions participating in the Gaia Multilateral Agreement.

(3) the \verb"Exoplanet Follow-up Observing Program" website, which is operated by the California Institute of Technology, under contract with the National Aeronautics and Space Administration under the Exoplanet Exploration Program.

(4) the Pan-STARRS1 surveys, which were made possible through contributions by the Institute for Astronomy, the University of Hawaii, the Pan-STARRS Project Office, the Max-Planck Society and its participating institutes, the Max Planck Institute for Astronomy, Heidelberg and the Max Planck Institute for Extraterrestrial Physics, Garching, The Johns Hopkins University, Durham University, the University of Edinburgh, the Queen's University Belfast, the Harvard-Smithsonian Center for Astrophysics, the Las Cumbres Observatory Global Telescope Network Incorporated, the National Central University of Taiwan, the Space Telescope Science Institute, and the National Aeronautics and Space Administration under Grant No. NNX08AR22G issued through the Planetary Science Division of the NASA Science Mission Directorate, the National Science Foundation Grant No. AST-1238877, the University of Maryland, Eotvos Lorand University (ELTE), and the Los Alamos National Laboratory. The Pan-STARRS1 Surveys are archived at the Space Telescope Science Institute (STScI) and can be accessed through MAST, the Mikulski Archive for Space Telescopes. Additional support for the Pan-STARRS1 public science archive is provided by the Gordon and Betty Moore Foundation.

(5) the SkyMapper survey, whose national facility capability has been funded through ARC LIEF grant LE130100104 from the Australian Research Council, awarded to the University of Sydney, the Australian National University, Swinburne University of Technology, the University of Queensland, the University of Western Australia, the University of Melbourne, Curtin University of Technology, Monash University and the Australian Astronomical Observatory. SkyMapper is owned and operated by The Australian National University's Research School of Astronomy and Astrophysics. The survey data were processed and provided by the SkyMapper Team at ANU. The SkyMapper node of the All-Sky Virtual Observatory (ASVO) is hosted at the National Computational Infrastructure (NCI). Development and support the SkyMapper node of the ASVO has been funded in part by Astronomy Australia Limited (AAL) and the Australian Government through the Commonwealth's Education Investment Fund (EIF) and National Collaborative Research Infrastructure Strategy (NCRIS), particularly the National eResearch Collaboration Tools and Resources (NeCTAR) and the Australian National Data Service Projects (ANDS). Full details of the SkyMapper DR1 data, processing, and early analysis are presented in \cite{wolf2018}.

\bibliography{mugrauer}

\begin{center}
\begin{table*}[h]
\caption{This table summarizes for all (C)TOIs (listed at first) and their detected co-moving companions their Gaia DR2 parallaxes $\pi$, proper motions $\mu$ in right ascension and declination, astrometric excess noises $epsi$, G-band magnitudes, as well as the used Apsis-Priam G-Band extinction estimates $A_{G}$ or if not available their G-Band extinctions, as listed either in the SHC (indicated with $\texttt{SHC}$) or derived from $A_V$ (indicated with $\maltese$).}\label{TAB_COMP_ASTROPHOTO}
\centering
\begin{tabular}{lccccccc}
\hline
TOI            & $\pi$              & $\mu_{\alpha}cos(\delta)$    & $\mu_{\delta}$ & $epsi$ & G     & $A_{G}$\\
               & [mas]              & [mas/yr]                     & [mas/yr]       & [mas]  & [mag] & [mag]\\
\hline
128\,A& $14.5754\pm0.0252$ & $-49.146\pm0.042$ & $45.307\pm0.046$ &0.000& $8.3717\pm0.0004$ & $0.0693_{-0.0547}^{+0.1308}$\\
128\,B& $14.5970\pm0.0228$ & $-49.450\pm0.044$ & $42.862\pm0.060$ &0.000& $10.9253\pm0.0019$ & $_{ }^{ }$\\
\hline
129\,A& $16.1560\pm0.0209$ & $-202.815\pm0.031$ & $-71.518\pm0.029$ &0.000& $10.5895\pm0.0004$ & $0.3230_{-0.1470}^{+0.1353}$\\
129\,B& $16.1171\pm0.0590$ & $-207.466\pm0.086$ & $-72.266\pm0.081$ &0.400& $15.3877\pm0.0008$ & $_{ }^{ }$\\
129\,C& $15.7568\pm0.2105$ & $-207.135\pm0.317$ & $-68.684\pm0.318$ &0.849& $18.6132\pm0.0019$ & $_{ }^{ }$\\
\hline
130\,A& $17.3368\pm0.0283$ & $139.424\pm0.056$ & $30.476\pm0.042$ &0.000& $7.8037\pm0.0003$ & $0.1095_{-0.1006}^{+0.1796}$\\
\hline
\end{tabular}
\end{table*}
\end{center}

\setcounter{table}{2}

\begin{table*} \caption{continued}
\begin{tabular}{lccccccccc}
\hline
TOI            & $\pi$              & $\mu_{\alpha}cos(\delta)$    & $\mu_{\delta}$ & $epsi$ & G     & $A_{G}$\\
               & [mas]              & [mas/yr]                     & [mas/yr]       & [mas]  & [mag] & [mag]\\
\hline
422\,B& $8.0032\pm0.0381$ & $-79.995\pm0.040$ & $-68.198\pm0.050$ &0.000& $9.3642\pm0.0002$ & $0.3548_{-0.1411}^{+0.1670}$\\
422\,A& $8.1757\pm0.0443$ & $-85.471\pm0.047$ & $-66.476\pm0.061$ &0.072& $8.2627\pm0.0004$ & $_{ }^{ }$\\
\hline
426\,A& $8.7748\pm0.0327$ & $5.456\pm0.052$ & $-25.589\pm0.062$ &0.000& $10.0409\pm0.0006$ & $ _{}^{}$\\
426\,B& $8.8513\pm0.0312$ & $4.352\pm0.047$ & $-27.448\pm0.051$ &0.000& $10.8149\pm0.0012$ & $ 0.0740_{-0.0660}^{+0.1900}$\\
\hline
444\,A& $17.3935\pm0.0249$ & $-25.830\pm0.035$ & $-116.428\pm0.045$ &0.000& $9.6063\pm0.0004$ & $0.0080_{-0.0027}^{+0.0250}$\\
444\,B& $17.3367\pm0.0648$ & $-24.508\pm0.093$ & $-118.363\pm0.116$ &0.440& $15.9063\pm0.0011$ & $_{ }^{ }$\\
\hline
\end{tabular}
\end{table*}

\setcounter{table}{2}

\begin{table*} \caption{continued}
\begin{tabular}{lccccccccc}
\hline
TOI            & $\pi$              & $\mu_{\alpha}cos(\delta)$    & $\mu_{\delta}$ & $epsi$ & G     & $A_{G}$\\
               & [mas]              & [mas/yr]                     & [mas/yr]       & [mas]  & [mag] & [mag]\\
\hline
756\,A& $11.5755\pm0.0419$ & $-216.372\pm0.066$ & $29.209\pm0.066$ &0.226& $13.6961\pm0.0005$ & $ _{}^{}$\\
756\,B& $11.5757\pm0.0641$ & $-215.513\pm0.109$ & $29.988\pm0.100$ &0.367& $15.0491\pm0.0005$ & $ 0.3130_{-0.1736}^{+0.4370}$\\
\hline
762\,A& $10.1292\pm0.0441$ & $-159.157\pm0.070$ & $-24.515\pm0.067$ &0.282& $14.9535\pm0.0005$ & $0.8440_{-0.4921}^{+0.5630}$\\
762\,B& $9.9166\pm0.1574$ & $-157.432\pm0.266$ & $-23.611\pm0.236$ &0.639& $18.2213\pm0.0017$ & $_{ }^{ }$\\
\hline
809\,A& $5.0986\pm0.0211$ & $8.506\pm0.042$ & $91.193\pm0.046$ &0.000& $10.6879\pm0.0003$ & $ _{}^{}$\\
809\,B& $5.1602\pm0.0588$ & $8.560\pm0.124$ & $91.211\pm0.128$ &0.290& $16.8111\pm0.0007$ & $ 0.1490_{-0.1220}^{+0.1640}$\\
\hline
\end{tabular}
\end{table*}

\setcounter{table}{2}

\begin{table*} \caption{continued}
\begin{tabular}{lccccccccc}
\hline
TOI            & $\pi$              & $\mu_{\alpha}cos(\delta)$    & $\mu_{\delta}$ & $epsi$ & G     & $A_{G}$\\
               & [mas]              & [mas/yr]                     & [mas/yr]       & [mas]  & [mag] & [mag]\\
\hline
1037\,A& $3.5472\pm0.0416$ & $-19.686\pm0.070$ & $10.953\pm0.068$ &0.137& $9.0671\pm0.0003$ & $0.2993_{-0.1754}^{+0.1437}$\\
1037\,B& $3.6947\pm0.1538$ & $-18.626\pm0.333$ & $10.334\pm0.278$ &0.737& $17.8010\pm0.0020$ & $_{ }^{ }$\\
\hline
1043\,A& $4.7777\pm0.0253$ & $-9.496\pm0.051$ & $14.659\pm0.045$ &0.000& $9.9162\pm0.0003$ & $0.7490_{-0.4327}^{+0.2856}$\\
1043\,B& $4.7792\pm0.2172$ & $-8.460\pm0.548$ & $14.309\pm0.400$ &0.899& $12.4824\pm0.0039$ & $_{ }^{ }$\\
\hline
1044\,A& $1.9937\pm0.0387$ & $-14.441\pm0.073$ & $6.214\pm0.069$ &0.062& $10.1590\pm0.0005$ & $_{}^{}$& \\
1044\,B& $3.1674\pm0.5553$ & $-13.558\pm0.943$ & $5.825\pm0.797$ &2.049& $10.7026\pm0.0010$ & $_{}^{}$\\
\hline
\end{tabular}
\end{table*}

\setcounter{table}{2}

\begin{table*} \caption{continued}
\begin{tabular}{lccccccccc}
\hline
TOI            & $\pi$              & $\mu_{\alpha}cos(\delta)$    & $\mu_{\delta}$ & $epsi$ & G     & $A_{G}$\\
               & [mas]              & [mas/yr]                     & [mas/yr]       & [mas]  & [mag] & [mag]\\
\hline
1242\,A& $9.0609\pm0.0298$ & $1.525\pm0.049$ & $-39.808\pm0.052$ &0.000& $12.3561\pm0.0002$ & $ _{}^{}$\\
1242\,B& $9.1904\pm0.0430$ & $1.614\pm0.086$ & $-39.539\pm0.089$ &0.267& $15.4298\pm0.0008$ & $ 0.1637_{-0.0717}^{+0.1703}$\\
\hline
1246\,A& $5.8736\pm0.0232$ & $-48.024\pm0.047$ & $81.928\pm0.051$ &0.000& $11.7291\pm0.0002$ & $0.1733_{-0.1489}^{+0.1717}$\\
1246\,B& $5.8666\pm0.0515$ & $-47.208\pm0.104$ & $81.950\pm0.115$ &0.321& $16.0599\pm0.0009$ & $_{ }^{ }$\\
\hline
1248\,A& $5.8976\pm0.0209$ & $-2.867\pm0.047$ & $-81.029\pm0.042$ &0.000& $11.6512\pm0.0002$ & $ _{}^{}$\\
1248\,B& $5.9462\pm0.0266$ & $-3.252\pm0.076$ & $-81.049\pm0.055$ &0.139& $14.9738\pm0.0006$ & $ 0.2808_{-0.0679}^{+0.4453}$\\
\hline
\end{tabular}
\end{table*}

\setcounter{table}{2}

\begin{table*} \caption{continued}
\begin{tabular}{lccccccccc}
\hline
TOI            & $\pi$              & $\mu_{\alpha}cos(\delta)$    & $\mu_{\delta}$ & $epsi$ & G     & $A_{G}$\\
               & [mas]              & [mas/yr]                     & [mas/yr]       & [mas]  & [mag] & [mag]\\
\hline
1516\,A& $4.0189\pm0.0221$ & $16.223\pm0.041$ & $-1.325\pm0.041$ &0.000& $10.7156\pm0.0002$ & $0.1320_{-0.1013}^{+0.1061}$\\
1516\,B& $3.9866\pm0.0850$ & $16.053\pm0.158$ & $-1.053\pm0.172$ &0.418& $17.2988\pm0.0028$ & $_{ }^{ }$\\
\hline
1520\,A& $5.0590\pm0.0260$ & $27.666\pm0.045$ & $8.250\pm0.035$ &0.000& $10.0906\pm0.0003$ & $0.6983_{-0.3974}^{+0.1917}$\\
1520\,B& $5.0657\pm0.0734$ & $29.643\pm0.122$ & $7.143\pm0.101$ &0.522& $16.4284\pm0.0008$ & $_{ }^{ }$\\
\hline
1521\,A& $3.1185\pm0.0258$ & $-6.510\pm0.044$ & $-11.243\pm0.042$ &0.000& $10.1862\pm0.0005$ & $ _{}^{}$\\
1521\,B& $3.1086\pm0.0273$ & $-6.403\pm0.052$ & $-11.656\pm0.044$ &0.000& $10.6684\pm0.0003$ & $ 0.9190_{-0.1744}^{+0.4836}$\\
\hline
\end{tabular}
\end{table*}

\setcounter{table}{2}

\begin{table*} \caption{continued}
\begin{tabular}{lccccccccc}
\hline
TOI            & $\pi$              & $\mu_{\alpha}cos(\delta)$    & $\mu_{\delta}$ & $epsi$ & G     & $A_{G}$\\
               & [mas]              & [mas/yr]                     & [mas/yr]       & [mas]  & [mag] & [mag]\\
\hline
1717\,A& $5.2886\pm0.0531$ & $1.214\pm0.053$ & $9.711\pm0.059$ &0.000& $9.8400\pm0.0003$ & $0.0988_{-0.0463}^{+0.1175}$\\
1717\,B& $5.3707\pm0.0411$ & $0.775\pm0.049$ & $8.590\pm0.046$ &0.192& $13.8273\pm0.0019$ & $_{ }^{ }$\\
\hline
1733\,A& $3.9514\pm0.0401$ & $-20.703\pm0.061$ & $-8.957\pm0.054$ &0.000& $10.8190\pm0.0004$ & $0.3383_{-0.2361}^{+0.2668}$\\
1733\,B& $4.1531\pm0.1481$ & $-19.368\pm0.185$ & $-9.298\pm0.179$ &0.763& $16.1581\pm0.0020$ & $_{ }^{ }$\\
1733\,C& $3.7297\pm0.2468$ & $-22.230\pm0.260$ & $-9.192\pm0.285$ &0.626& $17.6818\pm0.0028$ & $_{ }^{ }$\\
\hline
1744\,A& $8.7133\pm0.0315$ & $-1.548\pm0.058$ & $-18.707\pm0.049$ &0.000& $9.3918\pm0.0003$ & $0.0147_{-0.0103}^{+0.1494}$\\
\hline
\end{tabular}
\end{table*}

\setcounter{table}{2}

\begin{table*} \caption{continued}
\begin{tabular}{lccccccccc}
\hline
TOI            & $\pi$              & $\mu_{\alpha}cos(\delta)$    & $\mu_{\delta}$ & $epsi$ & G     & $A_{G}$\\
               & [mas]              & [mas/yr]                     & [mas/yr]       & [mas]  & [mag] & [mag]\\
\hline
1857\,A& $3.1209\pm0.0379$ & $6.779\pm0.053$ & $-1.198\pm0.048$ &0.084& $9.4651\pm0.0003$ & $0.9640_{-0.2254}^{+0.2170}$\\
1857\,B& $3.0542\pm0.0461$ & $7.955\pm0.066$ & $-1.420\pm0.059$ &0.153& $15.7271\pm0.0008$ & $_{ }^{ }$\\
\hline
1859\,A& $4.4278\pm0.0230$ & $12.120\pm0.044$ & $-12.156\pm0.041$ &0.000& $10.2843\pm0.0003$ & $0.2550_{-0.1136}^{+0.1354}$\\
1859\,B& $4.9684\pm0.5617$ & $15.763\pm0.904$ & $-13.259\pm1.003$ &0.360& $20.0989\pm0.0082$ & $_{ }^{ }$\\
\hline
1860\,A& $21.7747\pm0.0305$ & $-122.326\pm0.064$ & $110.588\pm0.058$ &0.000& $8.2322\pm0.0006$ & $0.0250_{-0.0190}^{+0.1048}$\\
1860\,BC& $21.8165\pm0.0995$ & $-127.622\pm0.191$ & $111.456\pm0.199$ &0.737& $14.7495\pm0.0013$ & $_{ }^{ }$\\
\hline
\end{tabular}
\flushleft{\underline{Comments:}\newline
$^1$: alias TOI 1113

$^2$: alias TOI 393

$^3$: alias TOI 1454

$^4$: alias TOI 1760

$^5$: is a white dwarf

}

\end{table*}

\setcounter{table}{2}

\begin{table*} \caption{continued}
\begin{tabular}{lccccccccc}
\hline
CTOI           & $\pi$              & $\mu_{\alpha}cos(\delta)$    & $\mu_{\delta}$ & $epsi$ & G     & $A_{G}$\\
               & [mas]              & [mas/yr]                     & [mas/yr]       & [mas]  & [mag] & [mag]\\
\hline
{\fontsize{7}{0}\selectfont 26547036\,A}& $9.1295\pm0.0267$ & $40.519\pm0.067$ & $11.106\pm0.066$ &0.000& $9.6957\pm0.0003$ & $0.1155_{-0.0811}^{+0.1181}$\\
{\fontsize{7}{0}\selectfont 26547036\,B}& $9.0701\pm0.1393$ & $38.904\pm0.594$ & $5.736\pm0.291$ &0.797& $14.3346\pm0.0144$ & $_{ }^{ }$\\
\hline
{\fontsize{7}{0}\selectfont 32150630\,A}& $6.5597\pm0.5590$ & $33.813\pm1.085$ & $33.631\pm1.113$ &4.424& $12.7356\pm0.0013$ & $1.1030_{-0.1261}^{+0.1160}$\\
{\fontsize{7}{0}\selectfont 32150630\,B}& $5.9566\pm0.0321$ & $34.080\pm0.064$ & $28.676\pm0.064$ &0.282& $14.7709\pm0.0004$ & $_{ }^{ }$\\
\hline
{\fontsize{7}{0}\selectfont 46029346\,A}& $2.7542\pm0.0255$ & $-8.353\pm0.049$ & $2.456\pm0.045$ &0.000& $12.2149\pm0.0014$ & $0.1720_{-0.0843}^{+0.0938}$\\
{\fontsize{7}{0}\selectfont 46029346\,B}& $3.0468\pm0.3023$ & $-8.042\pm0.558$ & $3.457\pm0.489$ &0.812& $19.2389\pm0.0026$ & $_{ }^{ }$\\
\hline
\end{tabular}
\end{table*}

\setcounter{table}{2}

\begin{table*} \caption{continued}
\begin{tabular}{lccccccccc}
\hline
CTOI            & $\pi$              & $\mu_{\alpha}cos(\delta)$    & $\mu_{\delta}$ & $epsi$ & G     & $A_{G}$\\
               & [mas]              & [mas/yr]                     & [mas/yr]       & [mas]  & [mag] & [mag]\\
\hline
{\fontsize{7}{0}\selectfont 233058265\,A}& $2.0187\pm0.0243$ & $-7.075\pm0.049$ & $-5.034\pm0.061$ &0.000& $12.1509\pm0.0002$ & $0.2468_{-0.1995}^{+0.1995}$& $\texttt{SHC}$\\
{\fontsize{7}{0}\selectfont 233058265\,B}& $1.8520\pm0.2105$ & $-7.383\pm0.447$ & $-5.307\pm0.624$ &0.801& $18.5794\pm0.0032$ & $_{}^{}$\\
\hline
{\fontsize{7}{0}\selectfont 233195052\,B}& $2.6521\pm0.0298$ & $-6.413\pm0.056$ & $18.876\pm0.059$ &0.006& $15.8211\pm0.0004$ & $ _{}^{}$\\
{\fontsize{7}{0}\selectfont 233195052\,A}& $2.7028\pm0.0145$ & $-6.463\pm0.027$ & $18.766\pm0.028$ &0.000& $13.7674\pm0.0003$ & $ 0.1195_{-0.0536}^{+0.2463}$\\
\hline
{\fontsize{7}{0}\selectfont 265489698\,A}& $4.7180\pm0.0284$ & $-27.259\pm0.036$ & $7.337\pm0.044$ &0.000& $9.1099\pm0.0017$ & $0.2169_{-0.2909}^{+0.2909}$& $\texttt{SHC}$\\
{\fontsize{7}{0}\selectfont 265489698\,B}& $4.8412\pm0.0289$ & $-28.097\pm0.046$ & $8.192\pm0.039$ &0.123& $13.8779\pm0.0041$ & $_{}^{}$\\
\hline
\end{tabular}
\end{table*}

\begin{table*} \caption{This table lists for each detected companion (sorted by its identifier) the angular separation $\rho$ and position angle $PA$ to the associated (C)TOI, the difference between its parallax and that of the (C)TOI $\Delta\pi$ with its significance (in brackets calculated by taking into account also the Gaia astrometric excess noise), the differential proper motion of the companion relative to the (C)TOI $\mu_{rel}$ with its significance, as well as its $cpm\text{-}index$. The last column indicates ($\bigstar$) if the detected companion is not listed in the WDS as companion(-candidate) of the (C)TOI.}\label{TAB_COMP_RELASTRO}
\begin{tabular}{lcccccccc}
\hline
TOI   & $\rho$    & $PA$        & $\Delta\pi$ & $sig$-      & $\mu_{rel}$ & $sig$-             & $cpm$-   & not in\\
             & [arcsec]  & [$^\circ$]  & [mas]       & $\Delta\pi$ & [mas/yr]    & $\mu_{rel}$ & $index$  & WDS\\
\hline
128\,B & $ 2.20038 \pm 0.00004 $ & $ 153.78053 \pm 0.00088 $ & $ 0.02 \pm 0.03 $ & 0.6 (0.6) & $ 2.46 \pm 0.08 $ & 32.7 & 54 &  \\
129\,B & $ 3.81144 \pm 0.00004 $ & $ 322.34105 \pm 0.00065 $ & $ 0.04 \pm 0.06 $ & 0.6 (0.1) & $ 4.71 \pm 0.09 $ & 51.6 & 92 &  \\
129\,C & $ 73.61906 \pm 0.00015 $ & $ 161.67257 \pm 0.00012 $ & $ 0.40 \pm 0.21 $ & 1.9 (0.5) & $ 5.17 \pm 0.32 $ & 16.2 & 84 & $\bigstar$ \\
130\,B & $ 2.25137 \pm 0.00027 $ & $ 280.82710 \pm 0.00662 $ & $ 0.46 \pm 0.26 $ & 1.8 (0.3) & $ 11.25 \pm 0.46 $ & 24.5 & 26 & $\bigstar$ \\
149\,B & $ 1.11156 \pm 0.00010 $ & $ 132.91564 \pm 0.00532 $ & $ 0.03 \pm 0.09 $ & 0.3 (0.1) & $ 0.06 \pm 0.27 $ & 0.2 & 754 &  \\
173\,B & $ 25.72579 \pm 0.00005 $ & $ 150.71133 \pm 0.00012 $ & $ 0.07 \pm 0.05 $ & 1.2 (0.3) & $ 0.87 \pm 0.10 $ & 8.5 & 70 & $\bigstar$ \\
\hline
\end{tabular}
\end{table*}

\setcounter{table}{3}

\begin{table*} \caption{continued}
\begin{tabular}{lccccccccc}
\hline
TOI   & $\rho$    & $PA$        & $\Delta\pi$ & $sig$-      & $\mu_{rel}$ & $sig$-             & $cpm$-   & not in\\
             & [arcsec]  & [$^\circ$]  & [mas]       & $\Delta\pi$ & [mas/yr]    & $\mu_{rel}$ & $index$  & WDS\\
\hline
686\,B & $ 12.95140 \pm 0.00041 $ & $ 30.97221 \pm 0.00163 $ & $ 0.79 \pm 0.40 $ & 2.0 (0.4) & $ 2.77 \pm 0.90 $ & 3.1 & 18 & $\bigstar$ \\
721\,B & $ 22.73395 \pm 0.00003 $ & $ 27.08926 \pm 0.00008 $ & $ 0.10 \pm 0.03 $ & 3.3 (0.9) & $ 2.18 \pm 0.06 $ & 37.8 & 55 &  \\
737\,B & $ 0.84011 \pm 0.00062 $ & $ 47.08799 \pm 0.04215 $ & $ 0.49 \pm 0.64 $ & 0.8 (0.3) & $ 3.94 \pm 1.45 $ & 2.7 & 29 & $\bigstar$ \\
756\,B & $ 11.09222 \pm 0.00005 $ & $ 163.19823 \pm 0.00035 $ & $ 0.00 \pm 0.08 $ & 0.0 (0.0) & $ 1.16 \pm 0.12 $ & 9.3 & 376 &  \\
762\,B & $ 3.19799 \pm 0.00012 $ & $ 239.54276 \pm 0.00210 $ & $ 0.21 \pm 0.16 $ & 1.3 (0.3) & $ 1.95 \pm 0.27 $ & 7.2 & 164 & $\bigstar$ \\
809\,B & $ 33.40695 \pm 0.00006 $ & $ 300.57194 \pm 0.00011 $ & $ 0.06 \pm 0.06 $ & 1.0 (0.2) & $ 0.06 \pm 0.13 $ & 0.4 & 3219 & $\bigstar$ \\
\hline
\end{tabular}
\end{table*}

\setcounter{table}{3}

\begin{table*} \caption{continued}
\begin{tabular}{lccccccccc}
\hline
TOI   & $\rho$    & $PA$        & $\Delta\pi$ & $sig$-      & $\mu_{rel}$ & $sig$-             & $cpm$-   & not in\\
             & [arcsec]  & [$^\circ$]  & [mas]       & $\Delta\pi$ & [mas/yr]    & $\mu_{rel}$ & $index$  & WDS\\
\hline
1220\,B & $ 7.22763 \pm 0.00006 $ & $ 131.71997 \pm 0.00048 $ & $ 0.09 \pm 0.07 $ & 1.4 (0.3) & $ 0.25 \pm 0.14 $ & 1.8 & 137 & $\bigstar$ \\
1228\,B & $ 11.54807 \pm 0.00004 $ & $ 70.14181 \pm 0.00019 $ & $ 0.06 \pm 0.04 $ & 1.6 (0.7) & $ 0.88 \pm 0.07 $ & 12.1 & 80 & $\bigstar$ \\
1234\,B & $ 2.16444 \pm 0.00007 $ & $ 121.97204 \pm 0.00210 $ & $ 0.09 \pm 0.11 $ & 0.8 (0.2) & $ 2.70 \pm 0.17 $ & 15.5 & 24 & $\bigstar$ \\
1242\,B & $ 4.43241 \pm 0.00005 $ & $ 353.06108 \pm 0.00069 $ & $ 0.13 \pm 0.05 $ & 2.5 (0.5) & $ 0.28 \pm 0.10 $ & 2.8 & 280 & $\bigstar$ \\
1246\,B & $ 4.51722 \pm 0.00005 $ & $ 71.42047 \pm 0.00073 $ & $ 0.01 \pm 0.06 $ & 0.1 (0.0) & $ 0.82 \pm 0.11 $ & 7.2 & 232 & $\bigstar$ \\
1248\,B & $ 31.26548 \pm 0.00003 $ & $ 10.58969 \pm 0.00007 $ & $ 0.05 \pm 0.03 $ & 1.4 (0.3) & $ 0.39 \pm 0.09 $ & 4.3 & 421 &  \\
\hline
\end{tabular}
\end{table*}

\setcounter{table}{3}

\begin{table*} \caption{continued}
\begin{tabular}{lccccccccc}
\hline
TOI   & $\rho$    & $PA$        & $\Delta\pi$ & $sig$-      & $\mu_{rel}$ & $sig$-             & $cpm$-   & not in\\
             & [arcsec]  & [$^\circ$]  & [mas]       & $\Delta\pi$ & [mas/yr]    & $\mu_{rel}$ & $index$  & WDS\\
\hline
1710\,B & $ 44.48009 \pm 0.00003 $ & $ 202.95690 \pm 0.00004 $ & $ 0.00 \pm 0.04 $ & 0.0 (0.0) & $ 0.30 \pm 0.07 $ & 4.3 & 543 &  \\
1715\,B & $ 4.97898 \pm 0.00046 $ & $ 191.57468 \pm 0.00546 $ & $ 0.03 \pm 0.55 $ & 0.0 (0.0) & $ 2.57 \pm 0.86 $ & 3.0 & 10 & $\bigstar$ \\
1717\,B & $ 2.87529 \pm 0.00005 $ & $ 183.05037 \pm 0.00072 $ & $ 0.08 \pm 0.07 $ & 1.2 (0.4) & $ 1.20 \pm 0.07 $ & 16.2 & 15 & $\bigstar$ \\
1733\,B & $ 29.46969 \pm 0.00010 $ & $ 57.12090 \pm 0.00019 $ & $ 0.20 \pm 0.15 $ & 1.3 (0.3) & $ 1.38 \pm 0.19 $ & 7.1 & 32 & $\bigstar$ \\
1733\,C & $ 30.20173 \pm 0.00016 $ & $ 58.03268 \pm 0.00031 $ & $ 0.22 \pm 0.25 $ & 0.9 (0.3) & $ 1.54 \pm 0.27 $ & 5.8 & 30 & $\bigstar$ \\
1744\,B & $ 3.57004 \pm 0.00005 $ & $ 148.85070 \pm 0.00070 $ & $ 0.06 \pm 0.05 $ & 1.3 (0.3) & $ 2.62 \pm 0.09 $ & 29.9 & 13 &   \\
\hline
\end{tabular}
\end{table*}

\setcounter{table}{3}

\begin{table*} \caption{continued}
\begin{tabular}{lccccccccc}
\hline
CTOI   & $\rho$    & $PA$        & $\Delta\pi$ & $sig$-      & $\mu_{rel}$ & $sig$-             & $cpm$-   & not in\\
             & [arcsec]  & [$^\circ$]  & [mas]       & $\Delta\pi$ & [mas/yr]    & $\mu_{rel}$ & $index$  & WDS\\
\hline
{\fontsize{7}{0}\selectfont 26547036\,B} & $ 1.91452 \pm 0.00015 $ & $ 32.99173 \pm 0.00548 $ & $ 0.06 \pm 0.14 $ & 0.4 (0.1) & $ 5.61 \pm 0.33 $ & 16.8 & 14 & $\bigstar$ \\
{\fontsize{7}{0}\selectfont 32150630\,B} & $ 11.09139 \pm 0.00055 $ & $ 147.19253 \pm 0.00286 $ & $ 0.60 \pm 0.56 $ & 1.1 (0.1) & $ 4.96 \pm 1.11 $ & 4.5 & 19 & $\bigstar$ \\
{\fontsize{7}{0}\selectfont 46029346\,B} & $ 20.80875 \pm 0.00027 $ & $ 156.07802 \pm 0.00071 $ & $ 0.29 \pm 0.30 $ & 1.0 (0.3) & $ 1.05 \pm 0.50 $ & 2.1 & 17 & $\bigstar$ \\
{\fontsize{7}{0}\selectfont 50380257\,B} & $ 10.40804 \pm 0.00004 $ & $ 160.06178 \pm 0.00019 $ & $ 0.01 \pm 0.04 $ & 0.3 (0.3) & $ 1.03 \pm 0.08 $ & 12.7 & 99 &  \\
{\fontsize{7}{0}\selectfont 53309262\,B} & $ 5.07856 \pm 0.00010 $ & $ 198.53382 \pm 0.00130 $ & $ 0.01 \pm 0.16 $ & 0.1 (0.0) & $ 0.75 \pm 0.18 $ & 4.2 & 99 & $\bigstar$ \\
{\fontsize{7}{0}\selectfont 94423678\,A} & $ 5.78481 \pm 0.00010 $ & $ 231.05611 \pm 0.00087 $ & $ 0.02 \pm 0.13 $ & 0.1 (0.0) & $ 1.06 \pm 0.16 $ & 6.7 & 54 & $\bigstar$ \\
\hline
\end{tabular}
\end{table*}

\begin{table*} \caption{This table lists the equatorial coordinates ($\alpha$, $\delta$ for epoch 2015.5) of all detected co-moving companions (sorted by their identifiers) together with their derived absolute G-band magnitudes $M_{G}$, masses, and effective temperatures $T_{eff}$. The flags for all companions, as defined in the text, are listed in the last column of this table.}\label{TAB_COMP_PROPS}
\begin{tabular}{lcccccccc}
\hline
TOI     & $\alpha$   & $\delta$   & $M_{G}$ & $sep$ & $mass$        &  $T_{eff}$ & Flags \\
& [$^\circ$] & [$^\circ$] & [mag]   & [au]  & [$M_{\odot}$] &  [K]       & \\
\hline
128\,B & 79.52265766889 & -80.96493231970 & $ 7.51 _{ -0.14 }^{ +0.06 }$ & 151 & $ 0.60 _{ -0.18 }^{ +0.19 }$&$ 4267 _{ -679 }^{ +1021 }$ &     \\
129\,B & 0.18448329645 & -54.82998468790 & $ 9.63 _{ -0.14 }^{ +0.15 }$ & 236 & $ 0.30 _{ -0.05 }^{ +0.01 }$&$ 3861 _{ -260 }^{ +183 }$ &   $\texttt{BPRP}$  \\
129\,C & 0.19677293298 & -54.85023527359 & $ 14.33 _{ -0.14 }^{ +0.15 }$ & 4557 & $ 0.10 _{ -0.01 }^{ +0.01 }$&$ 2709 _{ -18 }^{ +186 }$ &  $\texttt{inter}$ $\texttt{BPRP}$  \\
130\,B & 12.54114194511 & -83.74364962231 & $ 10.54 _{ -0.19 }^{ +0.11 }$ & 130 & $ 0.30 _{ -0.11 }^{ +0.15 }$&$ 3342 _{ -272 }^{ +315 }$ &     \\
149\,B & 76.63658503982 & -80.76821895578 & $ 3.17 _{ -0.09 }^{ +0.09 }$ & 428 & $ 1.39 _{ -0.04 }^{ +0.03 }$&$ 6632 _{ -74 }^{ +51 }$ &  $\texttt{inter}$ $\texttt{BPRP}$  \\
173\,B & 33.48022558031 & -80.58894310580 & $ 9.77 _{ -0.15 }^{ +0.03 }$ & 3928 & $ 0.42 _{ -0.08 }^{ +0.03 }$&$ 3340 _{ -31 }^{ +86 }$ &   $\texttt{BPRP}$ $\texttt{PRI}$ \\
\hline
\end{tabular}
\end{table*}

\setcounter{table}{4}

\begin{table*} \caption{continued}
\begin{tabular}{lccccccc}
\hline
TOI     & $\alpha$   & $\delta$   & $M_{G}$ & $sep$ & $mass$        &  $T_{eff}$ & Flags \\
& [$^\circ$] & [$^\circ$] & [mag]   & [au]  & [$M_{\odot}$] &  [K]       & \\
\hline
737\,B & 171.01497345473 & -19.55679993314 & $ 11.41 _{ -0.20 }^{ +0.20 }$ & 254 & $ 0.22 _{ -0.02 }^{ +0.02 }$&$ 3197 _{ -32 }^{ +22 }$ &  $\texttt{inter}$ \\
756\,B & 192.10500808405 & -45.47342057795 & $ 10.32 _{ -0.44 }^{ +0.18 }$ & 958 & $ 0.40 _{ -0.01 }^{ +0.01 }$&$ 3168 _{ -30 }^{ +10 }$ &   $\texttt{BPRP}$ $\texttt{PRI}$ \\
762\,B & 166.07360187839 & -47.82194669094 & $ 12.41 _{ -0.57 }^{ +0.50 }$ & 316 & $ 0.17 _{ -0.03 }^{ +0.03 }$&$ 3049 _{ -31 }^{ +80 }$ &  $\texttt{inter}$ $\texttt{BPRP}$  \\
809\,B & 82.96330580615 & -60.47200366911 & $ 10.44 _{ -0.17 }^{ +0.13 }$ & 6552 & $ 0.32 _{ -0.02 }^{ +0.01 }$&$ 3335 _{ -55 }^{ +26 }$ &   $\texttt{BPRP}$ $\texttt{PRI}$ \\
811\,B & 88.03026654527 & -32.92387115331 & $ 4.80 _{ -0.19 }^{ +0.14 }$ & 1241 & $ 0.92 _{ -0.03 }^{ +0.05 }$&$ 5516 _{ -133 }^{ +264 }$ &   $\texttt{BPRP}$ $\texttt{PRI}$ \\
815\,B & 155.87105216049 & -43.83337420960 & $ 8.37 _{ -0.30 }^{ +0.09 }$ & 361 & $ 0.55 _{ -0.01 }^{ +0.01 }$&$ 3835 _{ -52 }^{ +51 }$ &   $\texttt{BPRP}$ $\texttt{PRI}$ \\
\hline
\end{tabular}
\end{table*}

\setcounter{table}{4}

\begin{table*} \caption{continued}
\begin{tabular}{lccccccc}
\hline
TOI     & $\alpha$   & $\delta$   & $M_{G}$ & $sep$ & $mass$        &  $T_{eff}$ & Flags \\
& [$^\circ$] & [$^\circ$] & [mag]   & [au]  & [$M_{\odot}$] &  [K]       & \\
\hline
1234\,B & 290.57279878391 & 26.31917491824 & $ 11.73 _{ -0.16 }^{ +0.04 }$ & 228 & $ 0.20 _{ -0.01 }^{ +0.01 }$&$ 3114 _{ -64 }^{ +64 }$ &   $\texttt{BPRP}$  \\
1242\,B & 248.55299622090 & 60.19714896067 & $ 9.90 _{ -0.18 }^{ +0.08 }$ & 489 & $ 0.40 _{ -0.01 }^{ +0.01 }$&$ 3381 _{ -56 }^{ +13 }$ &   $\texttt{BPRP}$ $\texttt{PRI}$ \\
1246\,B & 251.11942105657 & 70.43039154298 & $ 9.73 _{ -0.18 }^{ +0.15 }$ & 769 & $ 0.39 _{ -0.02 }^{ +0.03 }$&$ 3488 _{ -37 }^{ +34 }$ &  $\texttt{inter}$ $\texttt{BPRP}$  \\
1248\,B & 259.02712070387 & 63.11414410059 & $ 8.09 _{ -0.45 }^{ +0.07 }$ & 5301 & $ 0.54 _{ -0.01 }^{ +0.01 }$&$ 3935 _{ -73 }^{ +21 }$ &   $\texttt{BPRP}$ $\texttt{PRI}$ \\
1253\,B & 260.90111371912 & 64.86363620801 & $ 8.14 _{ -0.74 }^{ +0.41 }$ & 992 & $ 0.60 _{ -0.01 }^{ +0.01 }$&$ 3842 _{ -23 }^{ +24 }$ &   $\texttt{BPRP}$ $\texttt{PRI}$ \\
1254\,B & 239.19850554781 & 65.88873486662 & $ 10.28 _{ -0.22 }^{ +0.13 }$ & 875 & $ 0.33 _{ -0.02 }^{ +0.03 }$&$ 3367 _{ -25 }^{ +42 }$ &  $\texttt{inter}$ $\texttt{BPRP}$  \\
\hline
\end{tabular}
\end{table*}

\setcounter{table}{4}

\begin{table*} \caption{continued}
\begin{tabular}{lccccccc}
\hline
TOI     & $\alpha$   & $\delta$   & $M_{G}$ & $sep$ & $mass$        &  $T_{eff}$ & Flags \\
& [$^\circ$] & [$^\circ$] & [mag]   & [au]  & [$M_{\odot}$] &  [K]       & \\
\hline
1717\,B & 104.62843451084 & 67.67653852758 & $ 8.65 _{ -0.12 }^{ +0.06 }$ & 544 & $ 0.50 _{ -0.18 }^{ +0.17 }$&$ 3822 _{ -415 }^{ +869 }$ &     \\
1733\,B & 118.89363263802 & 54.60104522918 & $ 9.11 _{ -0.27 }^{ +0.24 }$ & 7458 & $ 0.55 _{ -0.01 }^{ +0.01 }$&$ 3467 _{ -11 }^{ +39 }$ &   $\texttt{BPRP}$  \\
1733\,C & 118.89405099377 & 54.60104292681 & $ 11.17 _{ -0.27 }^{ +0.24 }$ & 7643 & $ 0.25 _{ -0.11 }^{ +0.11 }$&$ 3223 _{ -314 }^{ +272 }$ &     \\
1744\,B & 294.29212480683 & 60.88045326551 & $ 8.02 _{ -0.15 }^{ +0.02 }$ & 410 & $ 0.56 _{ -0.01 }^{ +0.03 }$&$ 3935 _{ -6 }^{ +55 }$ &  $\texttt{inter}$ $\texttt{BPRP}$  \\
1746\,B & 281.76777168782 & 54.91042776050 & $ 11.33 _{ -0.06 }^{ +0.06 }$ & 42 & $ 0.20 _{ -0.07 }^{ +0.15 }$&$ 3180 _{ -291 }^{ +281 }$ &     \\
1749\,B & 282.72824043706 & 64.41434948937 & $ 11.47 _{ -0.40 }^{ +0.14 }$ & 2308 & $ 0.25 _{ -0.01 }^{ +0.01 }$&$ 3105 _{ -3 }^{ +3 }$ &   $\texttt{BPRP}$  \\
\hline
\end{tabular}
\end{table*}

\setcounter{table}{4}

\begin{table*} \caption{continued}
\begin{tabular}{lccccccc}
\hline
CTOI     & $\alpha$   & $\delta$   & $M_{G}$ & $sep$ & $mass$        &  $T_{eff}$ & Flags \\
& [$^\circ$] & [$^\circ$] & [mag]   & [au]  & [$M_{\odot}$] &  [K]       & \\
\hline
{\fontsize{7}{0}\selectfont 26547036\,B} & 292.16742839087 & 53.48781540712 & $ 10.33 _{ -0.12 }^{ +0.09 }$ & 210 & $ 0.32 _{ -0.12 }^{ +0.14 }$&$ 3386 _{ -268 }^{ +344 }$ &     \\
{\fontsize{7}{0}\selectfont 32150630\,B} & 59.24915611605 & -68.62856241500 & $ 7.95 _{ -0.22 }^{ +0.23 }$ & 1691 & $ 0.60 _{ -0.02 }^{ +0.06 }$&$ 3716 _{ -89 }^{ +136 }$ &   $\texttt{BPRP}$ $\texttt{PRI}$ \\
{\fontsize{7}{0}\selectfont 46029346\,B} & 131.83080372881 & -57.19579870272 & $ 11.27 _{ -0.10 }^{ +0.09 }$ & 7555 & $ 0.23 _{ -0.01 }^{ +0.01 }$&$ 3212 _{ -10 }^{ +11 }$ &  $\texttt{inter}$ $\texttt{BPRP}$  \\
{\fontsize{7}{0}\selectfont 50380257\,B} & 32.98408606444 & -72.68149404570 & $ 3.71 _{ -0.16 }^{ +0.12 }$ & 1946 & $ 1.09 _{ -0.13 }^{ +0.16 }$&$ 6179 _{ -350 }^{ +444 }$ &   $\texttt{BPRP}$ $\texttt{PRI}$ \\
{\fontsize{7}{0}\selectfont 53309262\,B} & 116.97464432321 & 57.69711598929 & $ 11.64 _{ -0.38 }^{ +0.11 }$ & 625 & $ _{  }^{  }$&$  _{  }^{  }$ &  $\texttt{WD}$ $\texttt{BPRP}$  \\
{\fontsize{7}{0}\selectfont 94423678\,A} & 241.77040290903 & -30.55812881055 & $ 9.04 _{ -0.86 }^{ +0.42 }$ & 840 & $ 0.45 _{ -0.09 }^{ +0.05 }$&$ 3177 _{ -141 }^{ +37 }$ &   $\texttt{BPRP}$ $\texttt{PRI}$ \\
\hline
\end{tabular}
\end{table*}

\end{document}